\documentclass[fleqn,usenatbib]{mnras}

\usepackage{newtxtext,newtxmath}
\usepackage[T1]{fontenc}

\usepackage[normalem]{ulem} % WDW: PUT FOR USE OF STRIKETHROUGH
\DeclareRobustCommand{\VAN}[3]{#2}
\let\VANthebibliography\thebibliography
\def\thebibliography{\DeclareRobustCommand{\VAN}[3]{##3}\VANthebibliography}

\usepackage{graphicx}	% Including figure files
\usepackage{amsmath}	% Advanced maths commands
\title[]{First spatially resolved Na {\sc i} and He {\sc i} transitions towards an MYSO. Finding new tracers for the gaseous star/disc interface.}

\author[E. Koumpia et al.]{
Evgenia Koumpia,$^{1}$\thanks{E-mail: ekoumpia@eso.org}
M. Koutoulaki,$^{2}$ 
W.-J. de Wit,$^{1}$
R. D. Oudmaijer,$^{2}$
A. J. Frost,$^{3}$
S. L. Lumsden,$^{2}$
\newauthor and J. M. Pittard$^{2}$
\\
$^{1}$ESO, Alonso de Córdova 3107 Vitacura, Casilla, 19001, Santiago, Chile \\
$^{2}$School of Physics \& Astronomy, University of Leeds, Woodhouse Lane, LS2 9JT, Leeds, UK\\
$^{3}$Institute of Astronomy, KU Leuven, Celestijnlaan 200D, 3001, Leuven, Belgium}

\date{Accepted XXX. Received YYY; in original form ZZZ}

\pubyear{2022}

\begin{document}
\label{firstpage}
\pagerange{\pageref{firstpage}--\pageref{lastpage}}
\maketitle

\begin{abstract}
With steady observational advances, the formation of massive stars is being understood in more detail. Numerical models are converging on a scenario where accretion discs play a key role. Direct observational evidence of such discs at a few au scales is scarce, due to the rarity of such objects and the observational challenges, including the lack of adequate diagnostic lines in the near-IR. We present the analysis of K-band spectro-interferometric observations toward the Massive Young Stellar Object IRAS 13481-6124, which is known to host an accreting dusty disc. Using GRAVITY on the VLTI, we trace the crucial au-scales of the warm inner interface between the star and the accretion dusty disc. We detect and spatially resolve the Na {\sc i} doublet and He {\sc i} transitions towards an object of this class for the first time. The new observations in combination with our geometric models allowed us to probe the smallest -au- scales of accretion/ejection around an MYSO. We find that Na {\sc i} originates in the disc at smaller radii than the dust disc and is more compact than any of the other spatially resolved diagnostics (Br$\gamma$, He~{\sc i}, and CO). Our findings suggest that Na {\sc i} can be a new powerful diagnostic line in tracing the warm star/disc accreting interface of forming (massive) stars, while the similarities between He~{\sc i} and Br$\gamma$ point towards an accretion/ejection origin of He~{\sc i}.  
\end{abstract}

\begin{keywords}
stars: formation -- stars: massive -- accretion, accretion discs -- stars: individual: IRAS 13481-6124
\end{keywords}

%%%%%%%%%%%%%%%%%%%%%%%%%%%%%%%%%%%%%%%%%%%%%%%%%%

%%%%%%%%%%%%%%%%% BODY OF PAPER %%%%%%%%%%%%%%%%%%
\vspace{-0.2cm}
\section{Introduction}

Massive stars ($>$ 8 M$_\odot$) are among the most influential objects in space. Their birth, evolution, and death as supernovae highly contribute to the dynamical, chemical structure, and the evolution of their host galaxies. But what sets the conditions for the formation of a high-mass star? Our understanding of massive star formation is advancing due to new, high resolution observations \citep{Beltran2016}, yet some crucial issues remain unresolved. Theory suggests that massive stars can form according to the standard idea of accretion-ejection in discs \citep[e.g.,][]{Rosen2019,Klassen2016}, although more exotic mass assembly mechanisms cannot be discarded. Discs surrounding massive young stars are predicted to have relatively short lifetimes \citep[$<$ 10$^{5}$ yr; e.g.,][]{Kuiper2018}, likely as a result of stellar feedback leading to photoevaporation \citep[e.g.,][]{Owen2011,Ercolano2015}. Combined with their highly embedded nature and their large distances, detecting massive young star discs remains challenging.

Millimeter observations have revealed the first Keplerian-like disc structures around proto OB-type stars at large scales \citep[1000-1500 au;][]{Johnston2015,Ilee2016,Sanna2019,Zapata2019}, and at intermediate scales \citep[50-1000 au;][]{Maud2019,Motogi2019}. Direct observational evidence of accretion discs surrounding Massive Young Stellar Objects (MYSOs) at scales of a few au \citep{dewit2011,Kraus2017}, where the actual accretion occurs, is scarce \citep[e.g., IRAS 13481-6124 at 20 au;][]{Kraus2010,Caratti2020,Koumpia2021}. 

Near-IR emission lines are found to be excellent tracers of accretion and ejection processes in protostars. \citet{Ilee2013} performed detailed modelling of the CO bandhead emission towards a sample of 20 MYSOs, and found evidence of small-scale gaseous discs (few au up to tens of au) surrounding those objects \citep[see also,][]{Bik2004,Chandler1993,Chandler1995}. Practically all MYSOs display strong Br$\gamma$ line emission \citep{Bunn1995}, which had been attributed to the MYSO "disc-wind" scenario \citep{Drew1998}, or jet \citep{Caratti2016}, both important factors which affect the final mass of the central object. \citet{Pomohaci2017} report that about 37\% of MYSOs show the Na~{\sc i} doublet at 2.206 and 2.209 $\mu$m in emission, while 24\% show the He~{\sc i} transition. 

In the low-mass regime, \citet{Lyo2017} modelled kinematically the CO bandhead emission and the rare Na {\sc i} 2.2 $\mu$m doublet emission towards the low mass YSO ESO H-alpha 279a. Their results attribute Na {\sc i} to the inner region of a Keplerian rotating disc (0.04--1.00 au), while CO bandhead emission was found to originate somewhat further out (0.22 -- 3 au). Recently with ALMA, salt (NaCl) was detected towards young massive stars, uniquely probing both disc kinematics and physical conditions \citep{Ginsburg2019,Tanaka2020}, illustrating the general importance of Na~{\sc i} as a disc tracer. Therefore, Na {\sc i} appears to be a key tracer of the accretion/ejection process close to the protostar which subsequently affect its final mass and evolution. He {\sc i}, on the other hand, has been proposed as an accretion diagnostic towards Herbig Ae/Bes \citep{Oudmaijer2011}.

This work focuses on the MYSO IRAS\,13481-6124, which is one of the most massive (M$\sim$20~$M_{\odot}$), luminous (L$\sim$6.7$\times$10$^{4}$~$L_{\odot}$) MYSOs observable in the near-IR ($K=4^{m}.9$). IRAS 13481-6124 was the first MYSO to be imaged at milli-arcsecond scales with near-IR interferometry, and was found to host a hot disc \citep{Kraus2010}. The near-IR continuum emission is characterized by an elongated structure with a diameter size of $\sim$20~au at an inclination of $\sim$45$^{\circ}$. The presence of a bipolar molecular outflow, perpendicular to the disc and terminating in mid-IR bow-shocks, is also seen on scales of a few 10\,000 au. \citet{Caratti2016} report the first spatially resolved observations of the ionised gas content with near-IR interferometry and found that the Br$\gamma$ emission on au scales originates mainly from an ionised jet, while part of it appears to stem from a disc. Near-IR spectroscopic studies deliver strong evidence that the CO bandhead profile is kinematically consistent with a Keplerian rotating disc \citep{Ilee2013,Fedriani2020}. \citet{Caratti2020} presented the first spatially resolved CO bandhead emission, which was found to be well within the dusty rim around an MYSO. Importantly, the Na {\sc i} doublet in emission is also reported \citep{Pomohaci2017a,Fedriani2018}. In brief, IRAS\,13481-6124 has proven to be a key object with respect to spatially resolving tracers of the accretion process in MYSOs.  

We present new, more sensitive near-IR interferometric data of IRAS 13481-6124 using GRAVITY on the VLTI. We present the first spatially resolved observations of the Na {\sc i} doublet and He~{\sc i} transitions at mas scales, taken simultaneously with the spatially resolved CO bandhead, Br$\gamma$ and the hot dust emission at 2~$\mu$m. We constrain the emitting sizes using geometric models, and highlight the potential of Na {\sc i} and He {\sc i} for accretion studies in MYSOs.

\vspace{-0.3cm}

\section{Methods and observations}

\subsection{GRAVITY observations and data reduction}
\label{sec:obs} 

IRAS\,13481-6124 (MSX6C~G310.0135+00.3892; RA = 13$^{h}$51$^{m}$38$^{s}$, Dec = -61$^{\circ}$39\arcmin07\arcsec.5 [J2000]) was observed on the 23rd of March 2022 with GRAVITY \citep{Gravity_Coll2017,Eisenhauer2011} on the VLTI using the four 8.2-m Unit Telescopes (UTs). GRAVITY is a K-band interferometer which covers the spectral window between 1.99~$\mu$m and 2.45~$\mu$m. The observations were taken using the highest spectral resolution available (HR; R$\sim$4000) in combined polarisation mode, allowing to resolve spectral features down to $\sim$75~kms$^{-1}$. The observing run was performed under good atmospheric conditions with a seeing around 0.7\arcsec~and $\tau_{\rm cor}\sim6$~ms. No fringes could be secured on baselines involving UT4, but interferometric data of the source were obtained on the other three baselines.

%The observations were attempted twice on different nights but data recorded only on the second night.  Both observing attempts were done under ... The MYSO appeared diffuse and fainter than expected in the single UT images of the GRAVITY acquisition camera, resolving part of the MYSO K-band emission. As a result MACAO correction and injection into the GRAVITY fibres was compromised. 

The three projected baseline lengths are for UT12 43~m (PA$\sim$56$^{\circ}$), for UT23 50~m (PA$\sim$70$^{\circ}$) and for UT13 92~m (PA$\sim$64$^{\circ}$), resulting in an angular resolution as high as $\lambda/2B = 2.4$~mas at 2.2~$\mu$m. The data quality is not the same for each baseline, where UT2 was performing worse compared to UT1 and 3. At the distance of the source \citep[3.2~kpc; determined kinematically, see,][]{Lumsden2013} physical scales down to 7.7~au are resolved. The star HD 119073 (RA = 13$^{h}$42$^{m}$33$^{s}$, Dec = -57$^{\circ}$51\arcmin09\arcsec [J2000]) was observed to act as an interferometric calibrator. HD\,119073 subtends an angle of $\sim 0.4$~mas (i.e. spatially unresolved) and is characterised by a spectral type of K0III, and K-band magnitude of 5$^{m}$.4. 

The reduction and calibration of this dataset was performed using the standard GRAVITY pipeline recipes provided by ESO (version 1.5.4), delivering spectra of flux, visibility, differential phase and closure phase. The spectrum of the calibrator was used to correct the target spectrum for telluric features and instrumental response. %\footnote{The ESO pipeline for the GRAVITY data reduction can be downloaded at https://www.eso.org/sci/software/pipelines/gravity/ \rdo{these footnotes are taking a lot of space, candidates for removal if need be}})
\vspace{-0.4cm}
\subsection{Spectrum and visibilities}

The observed spectra of IRAS 13481-6124 cover the wavelength range between 2 and 2.4~$\mu$m. We report the detection of the following four atomic and molecular transitions: (1) hydrogen recombination emission at 2.167~$\mu$m (Br$\gamma$; Figure~\ref{fig:data1}), (2)  Na\,{\sc i} 2.206~$\mu$m and 2.209~$\mu$m doublet in emission (Figure~\ref{fig:data1}); (3) the CO bandhead in emission around 2.3-2.4~$\mu$m (Figure~\ref{fig:data3}); (4) He {\sc i} transition at 2.06~$\mu$m observed as a P-Cygni profile (Figure~\ref{fig:data3}). The P-Cygni profile seen along the Br$\gamma$ transition was previously reported in \citet{Caratti2016} and spectrally resolved in \citet{Fedriani2018}, revealing that the emission part consists of a narrow and a broad component.  
\vspace{-0.2cm}
\subsubsection{Hydrogen and Carbon-monoxide}

\begin{figure}
\begin{center}  
\includegraphics[scale=0.33]{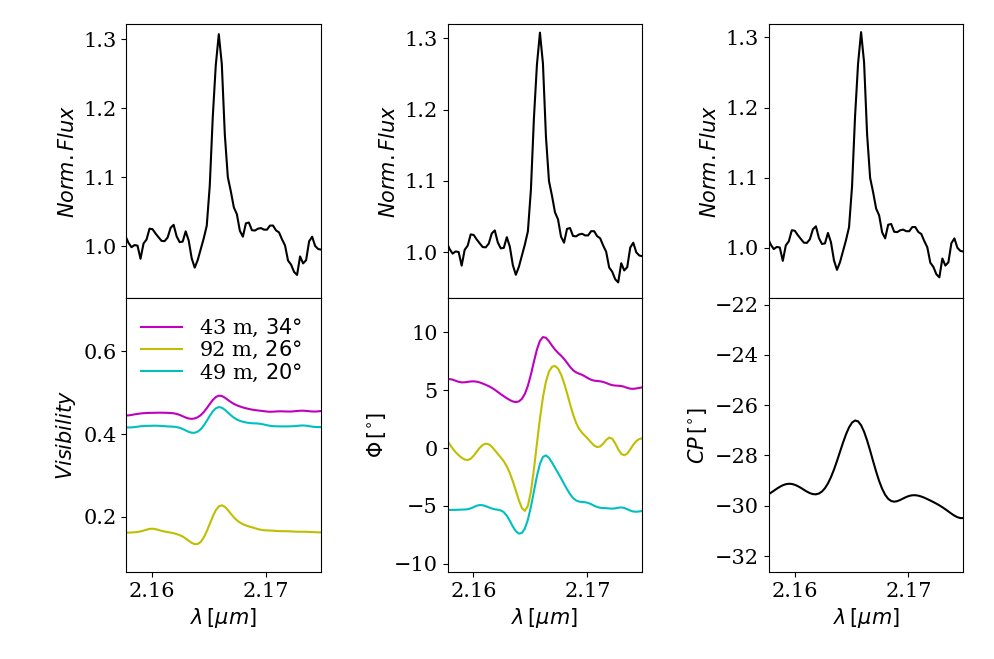} \\
\includegraphics[scale=0.30]{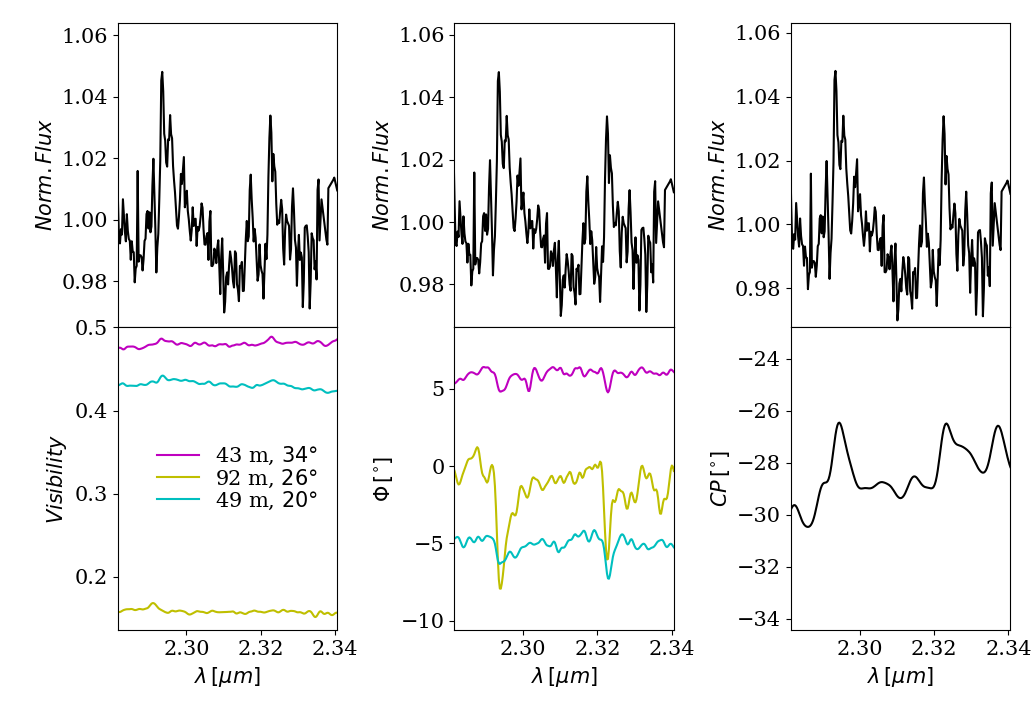} 
\end{center}
\caption{Relative flux, visibility, differential (with an offset of 5$^{\circ}$), and closure phase as a function of wavelength around the Br$\gamma$ emission (top) and CO (bottom) towards IRAS 13481-6124 using GRAVITY on the three UTs.}    
\label{fig:data1}
\end{figure}

\begin{figure}
\begin{center}  
\includegraphics[scale=0.32]{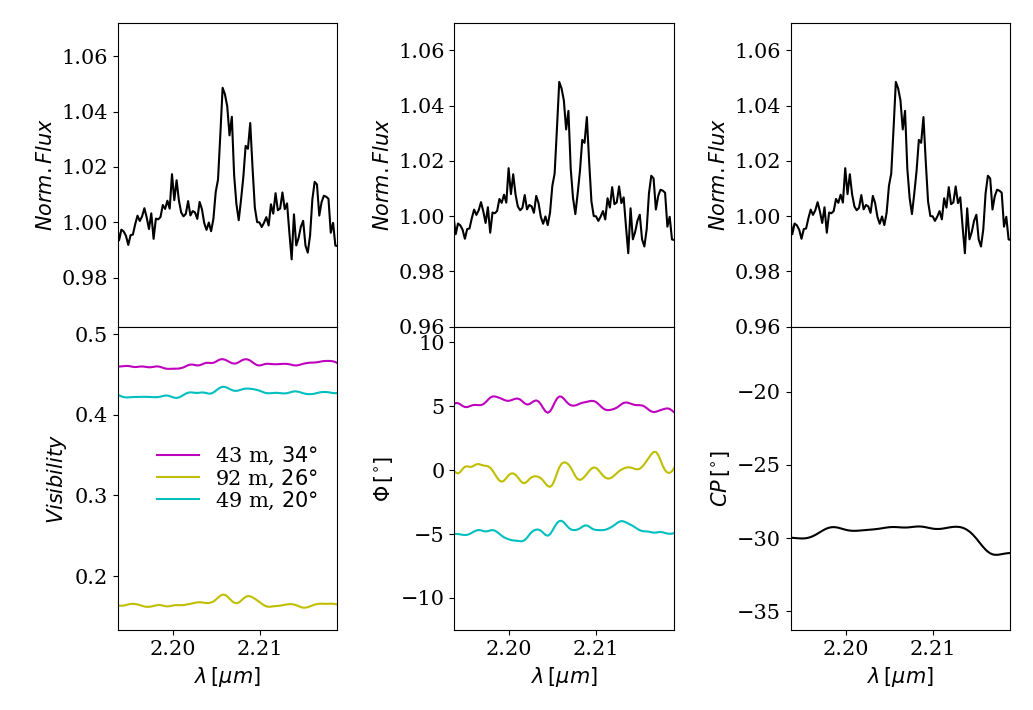} \\
\includegraphics[width=\columnwidth]{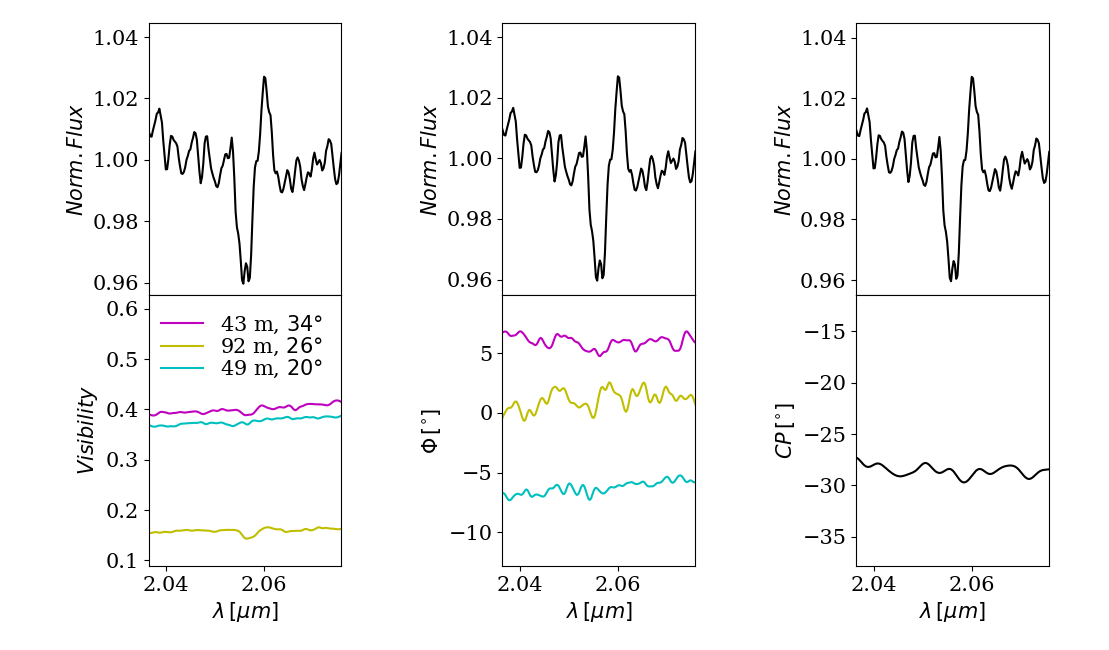} 
\end{center}
\caption{Same as Figure~\ref{fig:data1} but for Na {\sc i} (top) and He\,{\sc i} (bottom).}   
\vspace{-0.6cm}
\label{fig:data3}
\end{figure}

A visual inspection of the calibrated visibilities around the spectral lines, reveals some first information regarding the size of the emitting structures with respect to that of the continuum emission, the latter corresponding to the dust disc \citep{Kraus2010}. In Figures~\ref{fig:data1}-\ref{fig:data3} we focus on the visibilities of the Br$\gamma$, Na {\sc i}, He {\sc i} and the CO bandhead for the three baselines. The left column of the panels show the flux spectra around the line transitions of interest (top), with their corresponding calibrated visibilities (bottom).

The calibrated dispersed visibilities around Br$\gamma$ show a P-Cygni profile resembling the flux profile. It implies that the flux profile reflects a true P-Cygni and not a red-shifted emission component superposed on a stellar absorption spectrum. The visibility corresponding to the maximum line flux is larger than that of the continuum. This is indicative of a spatially resolved structure which is more compact than the continuum disc \citep[see also,][]{Caratti2016}. The absorption part of the visibility P-Cygni profile indicates that 
the blue-shifted emission originates from a structure larger than the continuum, and indeed larger than the average area of the ionised gas. This visibility signal can be interpreted as if Br$\gamma$ traces a bipolar jet/outflow, with the approaching component being larger than the component which moves away from the observer. The receding jet component is possibly shielded by the disc, as a result subtending a smaller angle on the sky than the approaching one.    

Looking at the calibrated visibilities produced by the CO bandhead emitting structure then we observe an increase in level, which is suggestive again of an emitting area smaller than that of the continuum disc. This is the second time CO bandhead emission is resolved in an MYSO \citep{Gravi_col_2020_b}.

\vspace{-0.2cm}
\subsubsection{Sodium and Helium}
New information on accretion in MYSOs is delivered by our GRAVITY observations that resolve spatially for the first time the structures emitting in the Na\,{\sc i} doublet and in the He\,{\sc i} transition. 

He\,{\sc i} visibility and flux profiles show a similar behaviour compared to those of Br$\gamma$. The peak-to-peak amplitude of the P-Cygni profile in both flux and calibrated visibilities are however smaller. Continuum corrected line visibilities are lower than those of Br$\gamma$. It implies that overall the He\,{\sc i} emission structure is the same as Br$\gamma$ albeit on average on a somewhat larger physical scale along the outflow/jet.

This is the first time that Na {\sc i} is reported to be spatially resolved, not only towards IRAS 13481-6124, but towards any YSO, both in the high-mass and low-mass regime. The observed increase in visibilities towards the wavelengths of the doublet is apparent and consistent in all three available baselines, and becomes more prominent in the longest baseline. Higher visibility values are indicative of a less spatially resolved emission. Therefore, our observations suggest that the Na {\sc i} doublet emitting area is smaller than the continuum and located in the interface between the star and the hot inner dusty rim.    

The qualitative inspection of the calibrated visibilities along the emission lines, provide us with a first picture of the distinct emitting areas. The emitting component of all observed transitions (He~{\sc i}, CO, Br$\gamma$, Na {\sc i}) appear to originate from the warm gaseous interface between the star and the innermost part of the dusty disc. The line profile of the visibility around Br$\gamma$ and He~{\sc i} suggests a bipolar outflow. The exact size of the individual gas components can only be derived via geometric modelling, which we present in Section~\ref{geom_sizes}. Nevertheless, the spatial information alone, hints that sodium and helium are new tracers of the innermost gaseous medium around (M)YSOs, increasing the number of suitable gas tracers in the NIR, which thus far has been limited to only Br$\gamma$ and CO.

%\begin{figure}
%\begin{center}  
%\includegraphics[scale=0.26]{G310_visibilities_all_lines.eps}
%\end{center}
%\caption{Normalised spectrum and visibilities around the Br$\gamma$, Na %{\sc i} and CO bandhead emission lines. {\bf These are not the final data - need re-working, tellurics, normalisation, binning etc, just checking the outline/format of the plot. SAME YRANGE for VIS plots. } }    
%\label{fig:data}
%\end{figure}
\vspace{-0.4cm}
\subsection{Phases}

In addition to the spectral information and the calibrated visibilities, GRAVITY provides information on the differential and closure phases in the entire spectral range. Based on the observed phases we provide some qualitative information on the distribution of the brightness distribution of each emission line observed in the spectrum (Figure~\ref{fig:data1}-\ref{fig:data3}), and in particular regarding the asymmetric or symmetric nature of the emission.  

The differential phases around the Br$\gamma$ emission show a characteristic P-Cygni profile (or else "S"-shape profile, similar to what is seen towards rotating discs), indicative of a different line displacement compared to the continuum depending if we look at the red-shifted or blue-shifted emission. Our findings are similar to those reported in \citet{Caratti2016}, and was further attributed to a collimated outflow. The detected increase in the single closure phase that could be obtained (a single triplet combination of the UTs), suggests a non-axisymmetric line brightness distribution. We note that the overall continuum emission is characterised by a closure phase of about -30$^{\circ}$, which is indicative of the asymmetric nature of the brightness distribution of the continuum too. On the other hand, the present observations around the Sodium emission do not show clear changes in differential or closure phases with respect to the continuum, which leads to the conclusion that this is a more symmetric emission region compared to the Br$\gamma$. A similar conclusion can be drawn regarding the symmetric nature of the He {\sc i} absorption; no significant changes are seen in differential or closure phases around this line. The situation is different when one inspects the differential and closure phases around the CO bandhead emission though. In that case we observe drops in the differential phases of up to $\sim$8$^{\circ}$ at the longest baseline (92~m), meaning that the CO emitting region is characerized by an offset of its photocentre compared to that of the continuum. The increase in the observed closure phase of about 3$^{\circ}$, suggests a non-axisymmetric emission. 
In conclusion, the continuum, Br$\gamma$ and CO bandhead emissions, all appear to have a non-axisymmetric brightness distribution, while the lines are characterised by an offset in their photocentre compared to the continuum emission. In contrast, both He {\sc i} and Na {\sc i} appear to be symmetric and to share the same photocentre with that of the continuum emission.  

\vspace{-0.7cm} 

\section{Geometric modelling and sizes}
\label{geom_sizes}

%The present GRAVITY observations contain visibility and differential phase measurements along the entire K-band spectrum (1.99~$\mu$m-2.45~$\mu$m) for three different baselines. The availability of only three baselines results in a single closure phase measurement for the observed spectral range. 
We determine the size of the continuum emission, and that of the distinct gas components as traced via the Na {\sc i}, Br$\gamma$, CO and He {\sc i}. To do so, we only use the measurements of the calibrated visibilities at the spectral channel where the emitting lines of interest peak, and the calibrated visibilities of the continuum at 2.2~$\mu$m. 

\begin{figure}
\includegraphics[scale=0.45]{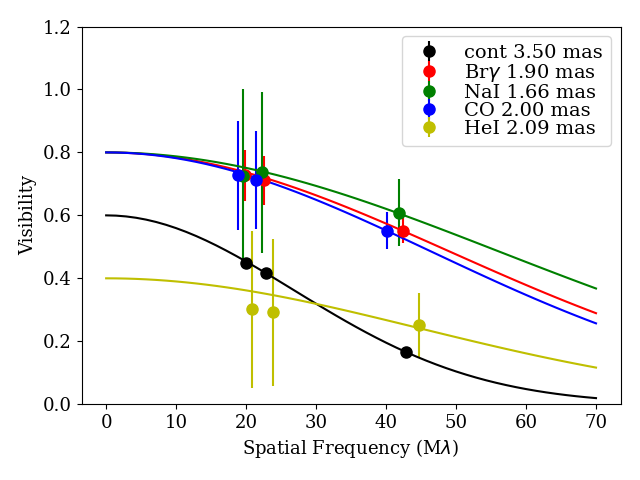}
\caption{Calibrated visibilities of the 2.2$\mu$m continuum emission, Na {\sc i} emission,  Br$\gamma$ emission, CO bandhead emission, and He {\sc i} emission towards IRAS 13481-6124 overplotted with the best fit geometrical Gaussian models.}    
\label{mod1}
\end{figure}

We fit the calibrated visibilities with simple geometrical models after adopting a Gaussian brightness distribution for a range of sizes (0.2~mas--10~mas) and taking into account any flux that is coming from diffuse emission (Figure~\ref{mod1}). The flux contribution from the diffuse emission was found to be 60\%, 40\%, and 20\% for the He {\sc i}, the continuum, and the rest of the lines (CO, Br$\gamma$, and Na {\sc i}) respectively. Figure~\ref{mod1} shows the calibrated visibilities extracted for each baseline length overplotted with the best-fit models for the individual emitting components. We find that the best-fit results in a size (diameter) of 3.50$\pm$0.30~mas (11~au) for the continuum, 1.66$\pm$0.20~mas (5.3~au) for the Na {\sc i}, 2.00$\pm$0.3~mas (6.4~au) for the CO, 1.90$\pm$0.40~mas (6~au) for Br$\gamma$, and 2.09$\pm$0.3~mas (6.7~au) for the He {\sc i}. 

To specify the origin of the 2.2$\mu$m continuum emission, we compare its measured size with that of the dust sublimation radius. We assume a lower and an upper temperature limit at which the dust is expected to sublimate \citep[1200-2000~K;][]{Kobayashi2011,Blasius2012}. Adopting the known luminosity of IRAS~13481-6124 of L$\sim$6.7$\times$10$^{4}$~$L_{\odot}$, we find that the dust sublimates at a distance of 5-15 au from the central star, which corresponds to 1.6-4.4~mas (or 3.2-8.8~mas; 10-30~au in diameter). The predicted range of dust sublimation radius is comparable to the measured size (11~au), and therefore we can safely conclude that the 2.2$\mu$m continuum emission originates from the hot dusty disc surrounding this MYSO. 

Our models reveal that the Na {\sc i}, He {\sc i}, Br$\gamma$ and CO bandhead emissions are all originating from gas interior to that of the dusty disc. The Na {\sc i} emission is located the closest to the star, and therefore, where active accretion takes place. The size of the CO emission is in perfect alignment with the inner radius of 2.8~au reported in \citet{Ilee2013}, which was based on kinematic modeling. Therefore, in this study we confirm spatially, what was previously determined based on spectral information. The compact nature of Na {\sc i} suggests that it acts as a tracer of the inner most regions of the YSO disc, from where up to now very little information could be extracted. On the other hand, He {\sc i}, which was previously proposed as an accretion diagnostic line towards the intermediate mass Herbig Ae/Bes \citep{Oudmaijer2011}, is found to be originating from an area somewhat larger but comparable to that of the Br$\gamma$ and CO bandhead emissions.  

\vspace{-0.2cm}

\section{Discussion}

\subsection{Na {\sc i} and He {\sc i}: two new tracers of the star/disc interface}

The atomic sodium doublet emission (Na {\sc i} at 2.20~$\mu$m) has been previously detected in various astrophysical environments \citep[e.g., evolved massive stars;][]{Hanson1996,Koumpia2022} including MYSOs \citep{Porter1998,Pomohaci2017}. Due to its low ionization potential (5.1~eV) and its co-existence in energetic environments with the hydrogen recombination line (Br$\gamma$), the physical process responsible for this emission has been under debate. Fluorescence (by means of 0.33 micron photon pumping) has been proposed as the most prominent scenario \citep{Scoville1983}. Another possible mechanism is that the Na {\sc i} emission is a result of collisional excitation in a dense medium \citep{Koumpia2020} where it can stay shielded from the direct stellar radiation (e.g., discs). Therefore, its exact location has been a long standing puzzle; it could originate from the accretion discs surrounding those newly forming stars. 

%The difficulty of constraining its exact location is a result of the observational limitations, and in particular the lack of adequate spectral resolution which could provide proper kinematic modelling of this emission \citep[similar to what has been previously done for the CO bandhead emission,][]{Ilee2013} and the high angular resolution needed to spatially resolve this emission. \rdo{is this sentence necessary? it reads as if you talk about our observations and it gives a glass-is-half-full feeling }

On the other hand, the Br$\gamma$ emission is a commonly used diagnostic line in young, forming stars, as not only is it abundantly present in those environments, but it is also often possible to spatially resolve it using NIR  spectro-interferometry. Recent studies suggest that for low-mass accreting protostars, Br$\gamma$ is located within the dust sublimation radius of a dusty disc and its location is consistent with magnetospheric accretion \citep[][]{Gravi_col_2020_b,Bouvier2020}. In hotter, more massive, stars, the Br$\gamma$ emission is found to be restricted to a quite compact area (1-10~au) within the dust sublimation radius, but larger than the region of influence of a magnetosphere in case of MHD accretion \citep[e.g.][]{Mendigutia2020,Garcia-Lopez2020}. The origin of the Br$\gamma$ emission around MYSOs has been contradictory in literature, with some studies attributing the emission to collimated jets/winds \citep[][]{Caratti2016,Davies2010}, or to disc wind and disc accretion models similar to Herbig Ae/Bes \citep{Koumpia2021}. The diversity seen in the literature in explaining the origin of the Br$\gamma$ emission is partly due to the intrinsic properties of this emission (i.e., high ionisation potential), which require conditions seen in the high energetic environments seen in both jets and innermost accreting matter, i.e., the observed emission could be a blend result of both processes. 

Detecting and identifying lines that can serve as disc accretion and ejection tracers in the NIR is of high importance. Here, we argue, that the Na {\sc i} doublet emission can be proved to be the new powerful diagnostic line in tracing the disc properties of the innermost warm disc around MYSOs (and YSOs). We present the first spatially resolved Na {\sc i} doublet emission around an MYSO (and in fact in any YSO environment independent of the central mass). Having a direct measurement of the exact location of  Na {\sc i} allows us to make a direct comparison with the CO bandhead at 2.3~$\mu$m, the origin of which has been attributed to the gaseous inner discs of MYSOs in the past. Based on medium-resolution (R $\sim$ 7000) NIR spectra of 36 MYSOs, the Na {\sc i} doublet emission appears co-present in all but one (M)YSOs where CO emission is present. Based on this, it has been argued that Na {\sc i} emission may require very similar physical conditions to those responsible for the CO bandhead emission \citep{Porter1998,Pomohaci2017,Lyo2017}, which typically stems from dense, warm conditions, as those seen in discs \citep{McGregor1988}. It is worth mentioning that all sources showing those lines in their spectrum are characterised by the presence of the Br$\gamma$ emission. 

A similar logic with the Br$\gamma$ emission applies for the He~{\sc i} transition. We spatially resolve it for the first time, but its complex spectral and visibility profiles, make it challenging to properly assess its underlying mechanism. Its similarity to the Br$\gamma$ emission points towards an ejection origin, but previous studies have associated it also with accretion around Herbig~Ae/Bes \citep{Oudmaijer2011}. A more detailed modeling of the full spectro-interferometric information is necessary to characterise the observed He~{\sc i}, but it is beyond the scope of this letter. Nevertheless, He {\sc i} can be proved to be an important diagnostic transition when studying the accretion / ejection processes, with a similar impact to that of the Br$\gamma$ emission.   

The very compact nature of the Na {\sc i} doublet emission, demonstrates that it arises from a very dense, warm, medium (where it gets shielded by the direct radiation), and that it has the potential to be a new key tracer of the warm inner disc surrounding forming stars, reaching scales even closer to the star compared to those of the CO bandhead. In addition to reaching the smallest scales in the accretion environment, Na {\sc i} can arguably serve as a more distinct disc tracer compared to the more energetic Br$\gamma$ and He {\sc i} which have been associated with both accretion and ejection processes in the past. 

\vspace{-0.8cm}

\section{Conclusions}

This paper presents new NIR high angular resolution interferometric observations (GRAVITY/VLTI) tracing material down to mas (au) scales of the MYSO IRAS 13481-6124. These observations probe simultaneously the neutral, ionised and molecular gaseous components located in the crucial star/disc interface around an MYSO. We report on the first spatially resolved Na {\sc i} doublet emission and He {\sc i} transition towards an object of this class (i.e., YSO).    

\begin{itemize}

\item The geometric models show that all gas components trace the interface between the star and the hot dust emission. In particular, the Na {\sc i} doublet line emission stems from a compact area which is located closest to the star (1.66~mas), while Br$\gamma$, CO bandhead, and the He {\sc i} are more extended but comparable to each other ($\sim$2~mas).

\item Our findings suggest that the Na {\sc i} doublet and the He {\sc i} transition are promising new tracers in the NIR of the innermost gaseous regions where accretion and ejection is the most active towards (M)YSOs. 

\item Br$\gamma$ and CO bandhead emissions, show changes in closure and differential phases which are characteristic of a non-axisymmetric brightness distribution, while implying an offset in the line emitting photocentre compared to that of the continuum. He {\sc i} and Na {\sc i} do not show significant changes in the observed phases, hinting at a symmetric brightness distribution while sharing the same photocentre with the continuum, at least at the traced scales and achieved sensitivity.

\end{itemize}

%Modelling the entire spectro-interferometric information, differential and closure phases is a scope of a follow-up paper (to reveal the specific geometries, asymmetries). The exact mechanism behind the relative location of the different gas components can be addressed by applying LTE models to fit the spectra together with the derived emitting sizes. This approach has been used successfully in the past to explain the origin of the Na {\sc i} from a smaller area compared to that of the Br$\gamma$ emission towards an object of a different Class \citep[YHG,][]{Koumpia2020}. 

Although discs are now detected, even the smallest, spatially resolved, structures are located well away (tenths of au) from the accretion regions closer to the star (sub-au to few au). To probe those, we need high spatial and spectral resolution in combination with emission lines that originate inside the dust sublimation region. Follow-up CRIRES+ observations will allow the kinematic modelling of the Na~{\sc i}, Br$\gamma$, He {\sc i} and the CO transitions, and trace any Keplerian-like motion and velocity gradient of the different gas components. Spectro-astrometry can be used to decouple disc and jet components. In this concept Na {\sc i} and He {\sc i} can be the new powerful diagnostic lines in tracing the gaseous star/disc accreting and ejecting interface, respectively, of forming (massive) stars. 

% Also trace kinematic gradient compared to Brgamma, CO. Study it in a larger sample (all sources that Pomohaci found to have NaI (observe them with GRAVITY+CRIRES+); will provide the unique opportunity to study the innermost parts of the disc via NaI and compare the disc properties with those of the central object (i.e., luminosity, evolution). Argue that CRIRES+ will allow us to confirm kinematically if NaI shows keplerian motion (when should we write a proposal)? 
\vspace{-0.8cm}
\section*{Acknowledgements}

Based on observations collected at the European Southern Observatory under ESO programme 0108.C-0693(D) (GRAVITY). We thank John Ilee for stimulating discussions.  

%%%%%%%%%%%%%%%%%%%%%%%%%%%%%%%%%%%%%%%%%%%%%%%%%%
\vspace{-0.6cm}
\section*{Data Availability} 
The reduced data presented in this paper will be shared on reasonable request to the corresponding author.


\vspace{-0.7cm}
\bibliographystyle{mnras}
\bibliography{example} % if your bibtex file is called example.bib

\bsp	% typesetting comment
\label{lastpage}
\end{document}